\documentclass[bbm,amsfonts,amsmath,amssymb,12pt]{article}
\usepackage[english]{babel}
\usepackage{amssymb,amsmath}
\usepackage{amscd}
\usepackage{color}

\usepackage{epsfig}
\usepackage{graphicx}
\usepackage{bm}
\usepackage{subfigure}
\begin{document}


\newcommand{\beq}{\begin{equation}}
\newcommand{\eeq}{\end{equation}}
\newcommand{\bea}{\begin{eqnarray}}
\newcommand{\eea}{\end{eqnarray}}
\newcommand{\beqn}{\begin{eqnarray}}
\newcommand{\eeqn}{\end{eqnarray}}
\newcommand{\beas}{\begin{eqnarray*}}
\newcommand{\eeas}{\end{eqnarray*}}
\newcommand{\defi}{\stackrel{\rm def}{=}}
\newcommand{\non}{\nonumber}
\newcommand{\bquo}{\begin{quote}}
\newcommand{\enqu}{\end{quote}}
\newcommand{\p}{\partial}


\def\de{\partial}
\def\Tr{ \hbox{\rm Tr}}
\def\const{\hbox {\rm const.}}
\def\o{\over}
\def\im{\hbox{\rm Im}}
\def\re{\hbox{\rm Re}}
\def\bra{\langle}\def\ket{\rangle}
\def\Arg{\hbox {\rm Arg}}
\def\Re{\hbox {\rm Re}}
\def\Im{\hbox {\rm Im}}
\def\diag{\hbox{\rm diag}}

\def\stroke{\vrule height8pt width0.4pt depth-0.1pt}
\def\topfleck{\vrule height8pt width0.5pt depth-5.9pt}
\def\botfleck{\vrule height2pt width0.5pt depth0.1pt}
\def\Zmath{\vcenter{\hbox{\numbers\rlap{\rlap{Z}\kern 0.8pt\topfleck}\kern
2.2pt\rlap Z\kern 6pt\botfleck\kern 1pt}}}
\def\Qmath{\vcenter{\hbox{\upright\rlap{\rlap{Q}\kern
3.8pt\stroke}\phantom{Q}}}}
\def\Nmath{\vcenter{\hbox{\upright\rlap{I}\kern 1.7pt N}}}
\def\Cmath{\vcenter{\hbox{\upright\rlap{\rlap{C}\kern
3.8pt\stroke}\phantom{C}}}}
\def\Rmath{\vcenter{\hbox{\upright\rlap{I}\kern 1.7pt R}}}
\def\Z{\ifmmode\Zmath\else$\Zmath$\fi}
\def\Q{\ifmmode\Qmath\else$\Qmath$\fi}
\def\N{\ifmmode\Nmath\else$\Nmath$\fi}
\def\C{\ifmmode\Cmath\else$\Cmath$\fi}
\def\R{\ifmmode\Rmath\else$\Rmath$\fi}


\def\QATOPD#1#2#3#4{{#3 \atopwithdelims#1#2 #4}}
\def\stackunder#1#2{\mathrel{\mathop{#2}\limits_{#1}}}
\def\stackreb#1#2{\mathrel{\mathop{#2}\limits_{#1}}}
\def\Tr{{\rm Tr}}
\def\res{{\rm res}}
\def\Bf#1{\mbox{\boldmath $#1$}}
\def\balpha{{\Bf\alpha}}
\def\bbeta{{\Bf\beta}}
\def\bgamma{{\Bf\gamma}}
\def\bnu{{\Bf\nu}}
\def\bmu{{\Bf\mu}}
\def\bphi{{\Bf\phi}}
\def\bPhi{{\Bf\Phi}}
\def\bomega{{\Bf\omega}}
\def\blambda{{\Bf\lambda}}
\def\brho{{\Bf\rho}}
\def\bsigma{{\bfit\sigma}}
\def\bxi{{\Bf\xi}}
\def\bbeta{{\Bf\eta}}
\def\d{\partial}
\def\der#1#2{\frac{\d{#1}}{\d{#2}}}
\def\Im{{\rm Im}}
\def\Re{{\rm Re}}
\def\rank{{\rm rank}}
\def\diag{{\rm diag}}
\def\2{{1\over 2}}
\def\ntwo{${\cal N}=2\;$}
\def\4N{${\cal N}=4$}
\def\none{${\cal N}=1\;$}
\def\x{\stackrel{\otimes}{,}}
\def\beq{\begin{equation}}
\def\eeq{\end{equation}}
\def\ba{\beq\new\begin{array}{c}}
\def\ea{\end{array}\eeq}
\def\be{\ba}
\def\ee{\ea}
\def\stackreb#1#2{\mathrel{\mathop{#2}\limits_{#1}}}

\def\baselinestretch{1.0}

\begin{titlepage}

\begin{flushright}
FTPI-MINN-05/39 \\
September 2005
\end{flushright}

\vspace{1cm}

\begin{center}

{\Large \bf Hidden Modulus in the
\\[6mm]  Extended Veneziano-Yankielowicz Theory}
\end{center}

\vspace{0.5cm}

\begin{center}
{Roberto {\sc Auzzi}$^{a}$ \quad and \quad Francesco {\sc Sannino}$^{b}$}
\end {center}
\begin{center}

$^a${\it  William I. Fine Theoretical Physics Institute,
University of Minnesota,
Minneapolis, MN 55455, USA}\\~~\\
$^b${\it The Niels Bohr Institute, Blegdamsvej 17, Copenhagen \O, Denmark.}\end{center}

\vspace{3mm}

\begin{abstract}
The issue of domain walls in the recently extended
Veneziano-Yankielowicz theory is investigated and we show that they
have an interesting substructure. We also demonstrate the presence
of a noncompact modulus. The associated family of degenerate
solutions is physically due to the presence of a valley of vacua in
the enlarged space of fields. This is a feature of the extended
Veneziano-Yankielowicz theory. Unfortunately the above properties do
not match the ones expected for the domain walls of $\mathcal{N}=1$
super Yang-Mills.
\end{abstract}

\end{titlepage}

\section{Introduction}
\label{intro}

Pure $\mathcal{N}=1$ $SU(N)$ super Yang-Mills theory
is known to possess $N$ independent vacua.
The issue of the domain walls between these vacua
is very interesting.
These objects are expected to saturate the central charge
of $\mathcal{N}=1$ superalgebra \cite{DS1} and the  tension
of the wall between the 1-st and the k-th vacua is \cite{DS2}:
 \[ T_k  \propto \sin \frac{\pi k}{N}. \]
 Moreover, there are arguments from string theory \cite{AV}
 and also from field theory \cite{RSV,ritz}
 which suggest that the walls have the following  multiplicity:
 \[ n_k=\frac{N !}{k ! (N-k)!}.\]

Unfortunately a dynamical description of these domain walls using
the basic fields of super Yang-Mills or the composite ones is still
lacking. In the framework of the Veneziano-Yankielowicz (VY)
effective lagrangian it is not possible to find a wall solution
interpolating the different vacua \cite{ks,kss,cs,kks,smilga}. We
remind the reader that the VY theory concisely summarizes the
symmetry of the underlying theory in terms of the following
composite chiral superfield
\begin{eqnarray}
S=\frac{3}{32\pi^2\, N}{\rm Tr}W^2 \ ,
\end{eqnarray}
where $W_{\alpha}$ is the supersymmetric field strength. When
interpreting $S$ as an elementary field it describes a gluinoball
and its associated fermionic partner.

The K\"ahler potential and the superpotential are respectively \cite{VY}
\begin{eqnarray}
K(S,\bar S)=\frac{9N^2}\alpha(S\bar S)^{\frac13}\,,\qquad
W_{VY}[S]=\frac{2N}{3}S\log_{(0)}
\left(\frac{S}{e\Lambda^3}\right)^N\,,
\end{eqnarray}
where $\alpha$ is a dimensionless real parameter and we restricted ourselves to the first branch of the logarithm.
The complex bosonic degree of freedom $\phi_s$ in the VY Lagrangian posses the same quantum numbers of
the gaugino bilinear $\lambda \lambda$. The latter is expected to condense spontaneously breaking the $Z_{2N}$ symmetry to $Z_2$ and
leaving behind $N$ vacua.

On general grounds there are in principle two possibilities for the domain walls; the wall can pass in the real $S$ space
through the value $\phi_s=0$ or it can pass away from this value
in the complex $\phi_s$ plane.
The real wall solution is somehow a very particular one: it can interpolate
only between two vacua aligned with the origin. This implies that we can have a real wall passing through the origin only when $N$ is even and $k=N/2$. Then
there will be $N/2$ pairs of vacua aligned with the origin. In the VY Lagrangian approach there are problems
for both the possibilities: the real wall solution stops at the origin at $\phi_s=0$
and so cannot interpolate between the two vacua at $\phi_s=\pm \phi_{s0}$; for the complex
wall there are problems due to the logarithmic cuts in the superpotential.

It is natural at this point to study the domain wall problem in the recently proposed extension of the
VY effective Lagrangian \cite{EVY1}. Real domain walls
solutions were found in this approach in Ref. \cite{EVY3}; in this work
we further investigate these domain wall solutions.

We will study the two color theory, however our results are
generally applicable to an even number of colors for the domain wall
solutions interpolating between two vacua aligned with the origin.
We will demonstrate that the domain wall solutions possess a hidden
noncompact modulus. The presence of such a modulus is due to the
fact that there exists a valley of vacua at the origin in the
enlarged space of fields with respect to the original VY theory
which contains only one chiral superfield.

In the next section we briefly introduce the extended VY theory while in section three we show how a modulus emerges
when investigating the real domain walls in this theory.
 In section four we briefly discuss the issue of complex walls.
  We finally conclude in section five and provide also some outlook.

\section{Extended VY Theory}
There are many reasons to consider extending the VY theory. When
trying to break SYM to YM one would expect the glueball degrees of
freedom to be present, for example. However in the original theory
such states are missing.   Many different approaches were considered
in the literature to extend the VY theory and the reader can find an
exhaustive list of references in \cite{EVY1}. In \cite{EVY1} a
number of different consistency checks was shown to naturally lead
to the following form of the extended VY superpotential: \beq
W(S,\chi)=\frac{2 N^2}{3} S \left[ \log_{(0)} S -1 -\log \left( -e
\frac{\chi}{N} \log_{(0)} \chi^N \right) \right]. \eeq A K\"{a}hler
potential is needed to investigate various dynamical properties. The
simplest one for $S$ which is consistent with  the quantum
anomalies in the VY theory is $(S \bar{S})^{1/3}$. Due to the
presence of the new field $\chi$ one can modify the K\"{a}hler potential in
order to provide a kinetic term also for $\chi$: \beq
K(S,\bar{S},\chi,\bar{\chi})=\frac{9 N^2}{\alpha} (S \bar{S})^{1/3}
h(\chi,\bar{\chi}), \label{kahler} \eeq where $h(\chi,\bar{\chi})$
is a real positive definite function. This is the generic form of
the K\"{a}hler potential in order not to upset the saturation of quantum
anomalies. The associated K\"{a}hler metric is: \beq g_{i
\bar{j}}=\frac{\partial^2 K}{\partial \phi^i \partial
\bar{\phi}^{\bar{j}}}. \eeq The potential is: \beq V=\frac{\partial
W}{\partial \phi^i} g^{i \bar{j}} \frac{\partial \bar{W}}{\partial
\bar{\phi}^{\bar{j}}}, \eeq where \beq g^{i \bar{j}}=(g_{i
\bar{j}}^{-1})^T. \eeq In reference \cite{EVY2} it was argued that
the glueball states are heavier than the gluinoball states. If this
were not the case there would have been no reason to saturate the
underlying anomalies of the theory using only the composite objects
constructed out of the gluinoball fields. Since the extended VY
theory well describes the vacuum properties of SYM and the spectrum
of the lowest lying states of the theory one would hope that it can
also overcome the problems related to the domain wall solutions and
their multiplicity.

\section{Real BPS walls: A new moduli space}
Recently the domain walls in the extended VY for $N=2$ were investigated
in Ref. \cite{EVY3}, where real domain wall solutions were found to exist.
We recall the general BPS domain wall equations for a
generic supersymmetric Wess-Zumino model with more than one field labeled by Latin letters:
\beq \frac{d \phi^i}{d z}=e^{-i \beta} g^{i \bar{j}} \partial_{\bar{j}} \bar{W}. \eeq
We have only two scalar fields $\phi^1=\phi_S$ and $\phi^2=\phi_\chi$.
The system of differential equations reduces to (see Ref. \cite{EVY3})
\begin{eqnarray}
\frac{\partial}{\partial z}\left(%
\begin{array}{c}
   \varphi\\
  \varphi_{\chi} \\
\end{array}%
\right) = \frac{\alpha}{N^2} \frac{(\bar{\varphi}\varphi)^{\frac{2}{3}}}{h\, h_{\chi\bar{\chi}} - h_{\chi}\bar{h}_{\bar{\chi}}} \left[%
\begin{array}{cc}
  h_{\chi\bar{\chi}} & -\frac{{\bar{\varphi}^{-1}}h_{\chi}}{3} \\
  -\frac{\varphi^{-1}\bar{h}_{\bar{\chi}}}{3} & \frac{h}{9}  ({\bar{\varphi}\varphi})^{-1}\\
\end{array}%
\right] \,\left(%
\begin{array}{c}
  \partial_{\bar{S}}{\overline{W}} \\
   \partial_{\bar{\chi}}{\overline{W}}\\
\end{array}%
\right)\,, \label{DE}
\end{eqnarray}
with $h_{\chi} = \partial_{\chi}h$, $\bar{h}_{\bar{\chi}}=\partial_{\bar{\chi}}h$ and $h_{\chi\bar{\chi}}= \partial_{\chi}\partial_{\bar{\chi}}h$.
 We also have
\begin{eqnarray}
 \partial_{{S}}{{W}}=\frac{2N^2}{3}\,\log_{(0)}
\left(\frac{S}{-e\frac{\chi}{N} \log_{(0)} \chi^N}\right)\ ,\qquad  \frac{\partial_{\chi}{W}}{S}=\frac{2N}{3} \, \frac{\log_{(0)} \chi^N + N}{-\frac{\chi}{N}\log_{(0)}  \chi^N}\,.
\end{eqnarray}

\begin{figure}[h]
\begin{center}
\epsfxsize=2in \epsffile{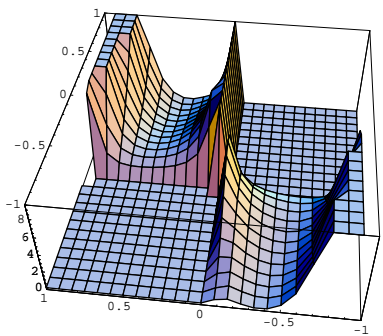}
\end{center}
\caption{\footnotesize Potential $V(\phi_s,\phi_\chi)$ for $N=2$ and for
 the K\"{a}hler $h_2$ for real $\phi_s$, $\phi_\chi$.
It goes to infinity at $\phi_\chi=0$ and at $\phi_\chi=\pm 1$ ;
 there is a valley of zero potential at $\phi_s=0$.}
\label{fig1}
\end{figure}
The potential $V$
for real $\phi_s,\phi_\chi$ can be written in the following way:
\beq V(\phi_s,\phi_\chi)= |\phi_s|^{4/3} (f(\phi_\chi) \log |\phi_s|^2+g(\phi_\chi))\eeq
where $f,g$ are two functions of $\phi_\chi$ depending on
 $h(\chi,\bar{\chi})$ in Eq. (\ref{kahler}).
The potential $V$ has a valley of zeros at $\phi_s=0$. This valley
of vacua exists for every function $h$ in Eq. (\ref{kahler}) and so
it is a general property of the extended VY Lagrangian (see Fig.
\ref{fig1}). Actually the point $\phi_s=\phi_\chi=0$ is not part of
this valley of vacua. The potential assumes different values
according to the path one uses when approaching the origin.

The existence and behavior of the walls depend not only on the
superpotential of the effective theory, but also on the K\"{a}hler potential.
In Ref. \cite{EVY3} the following K\"{a}hler potentials were
considered and motivated: \beq h_1(\chi,\bar{\chi})=(1+\gamma e^2
\chi \bar{\chi}) (e^2 \frac{\chi \bar{\chi}}{N^2} \log{\chi^N}
\log{\bar{\chi}^N})^{-\frac{1}{3}}, \eeq and \beq
h_2(\chi,\bar{\chi})=(1+\gamma e^2 \chi \bar{\chi}) (e^2 \chi
\bar{\chi})^{-\frac{1}{3}}. \eeq Explicit solutions interpolating
two opposite vacua $(\phi_s, \phi_\chi)= (\pm \phi_{S0},\pm
\phi_{\chi 0})$ were found for both the K\"{a}hler potentials. These
solutions pass through the origin in the field space $(\phi_s,
\phi_\chi)=(0,0)$. We are going to present a detailed analysis for
the theory assuming the K\"{a}hler potential $h_2$; the case in which one uses
$h_1$ is analogous. The reason is that the two K\"{a}hlers have a
similar behavior near the origin $\phi_s,\phi_\chi=0$.

\begin{figure}[h]
\begin{center}
$\begin{array}{c@{\hspace{.2in}}c@{\hspace{.2in}}c} \epsfxsize=1.5in
\epsffile{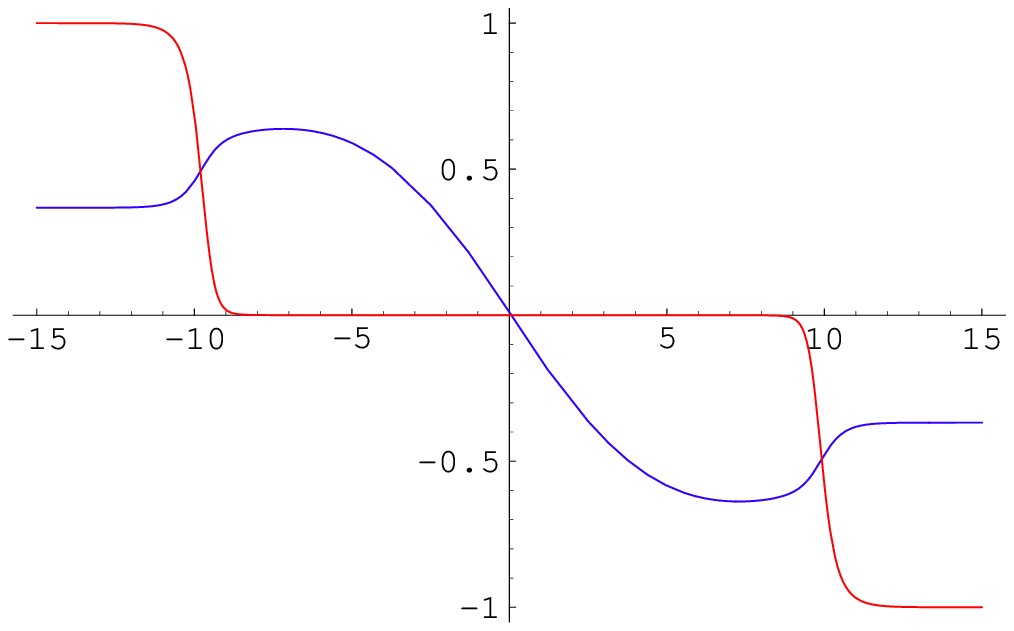} &
    \epsfxsize=1.5in
    \epsffile{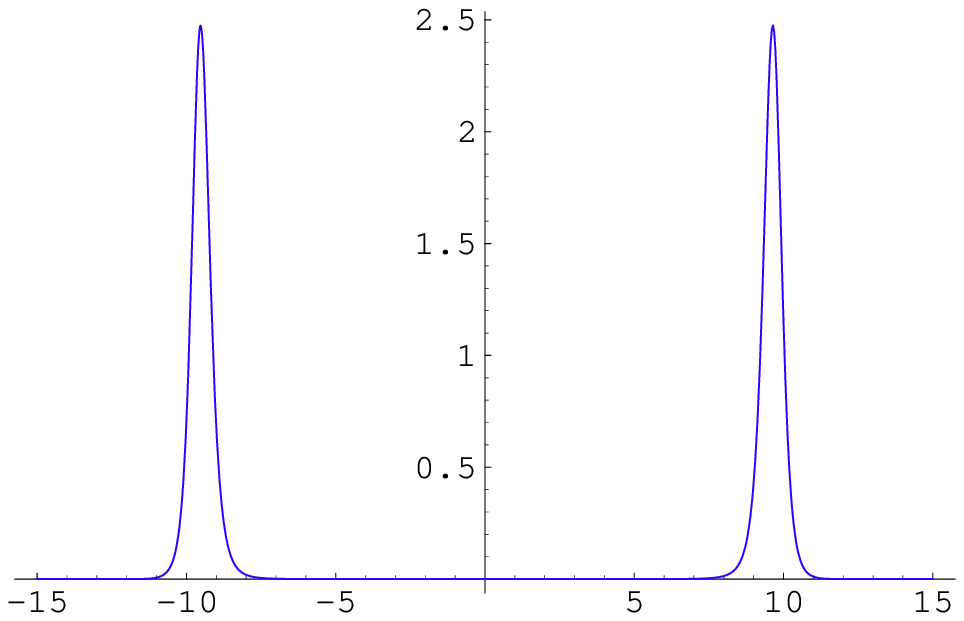} &
     \epsfxsize=1.5in
    \epsffile{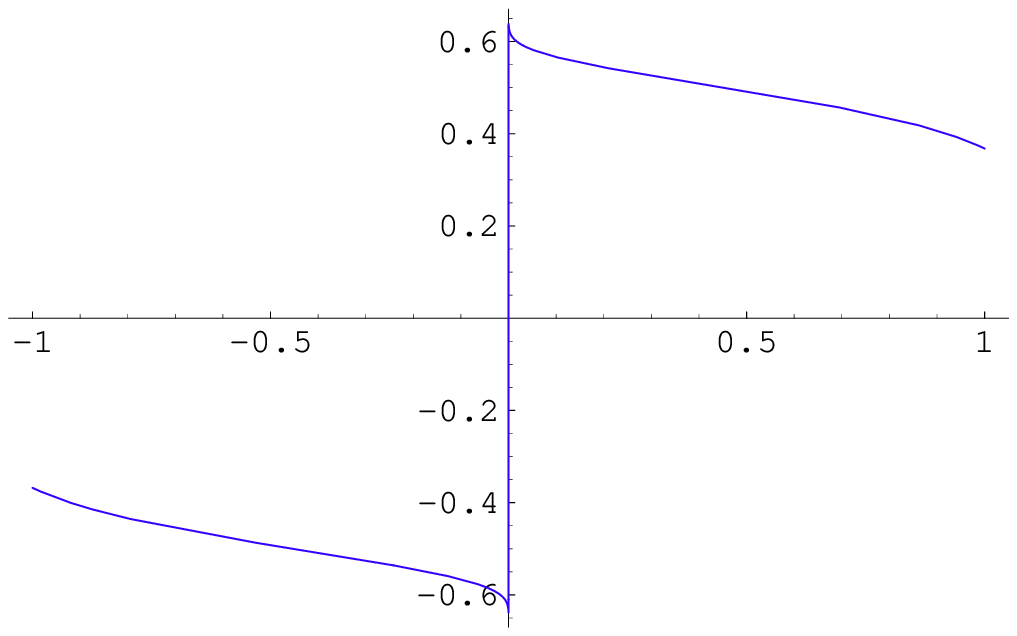}
\end{array}$
\end{center}
\caption{\footnotesize Wall for the K\"{a}hler potential $h_2$ and
at large $l$. Left: Field profiles for $\phi_s(z)$ and
$\phi_\chi(z)$. Center: Energy density. Right: Field profiles in the
plane $(\phi_s,\phi_\chi)$. The fields pass through the valley at
$\phi_s=0$.} \label{fig2}
\end{figure}

We have tried  to generalize the analysis to the following class of
K\"{a}hler potentials: \beq h(\chi,\bar{\chi})=(1+\gamma e^2 \chi
\bar{\chi}) (e^2 \chi \bar{\chi})^{p}. \eeq If we keep only the
dominant terms near the origin we find the following equation: \beq
\frac{d \phi_s}{d \phi_\chi}=-3 p \frac{\phi_s}{\phi_\chi},\eeq
which has a regular solution near the origin only if $-3 p$ is a
positive integer: \beq \phi_s=C \phi_\chi^{-3 p}. \label{mocambo}
\eeq We have explored different values of $p$. However it turns out
that the only acceptable value, in order to have a well behaved
domain wall solution, is $p=-1/3$.

\begin{figure}[h]
\begin{center}
$\begin{array}{c@{\hspace{.2in}}c@{\hspace{.2in}}c} \epsfxsize=1.5in
\epsffile{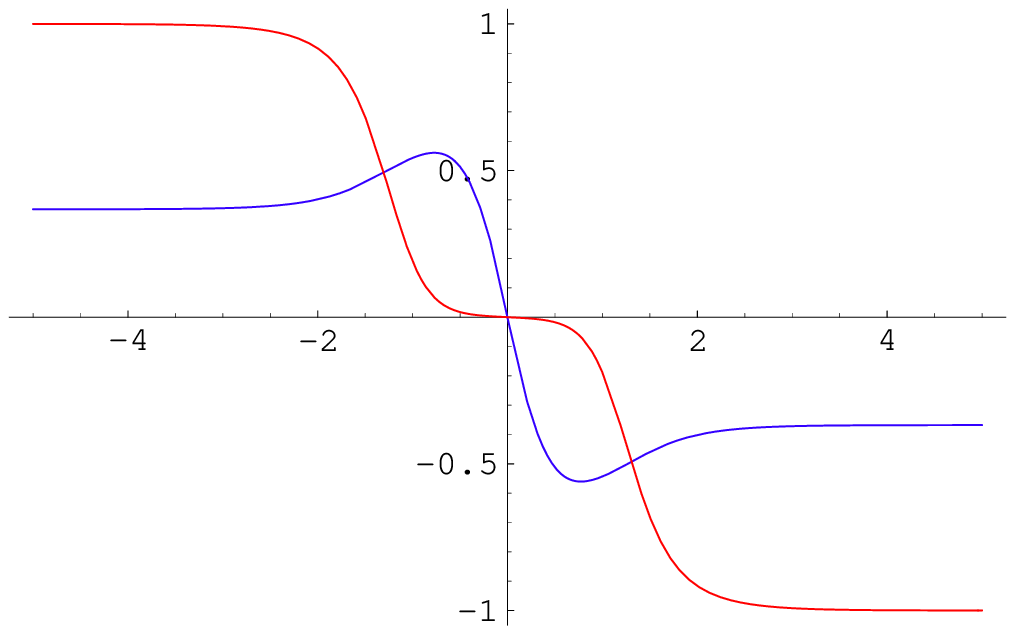} &
    \epsfxsize=1.5in
    \epsffile{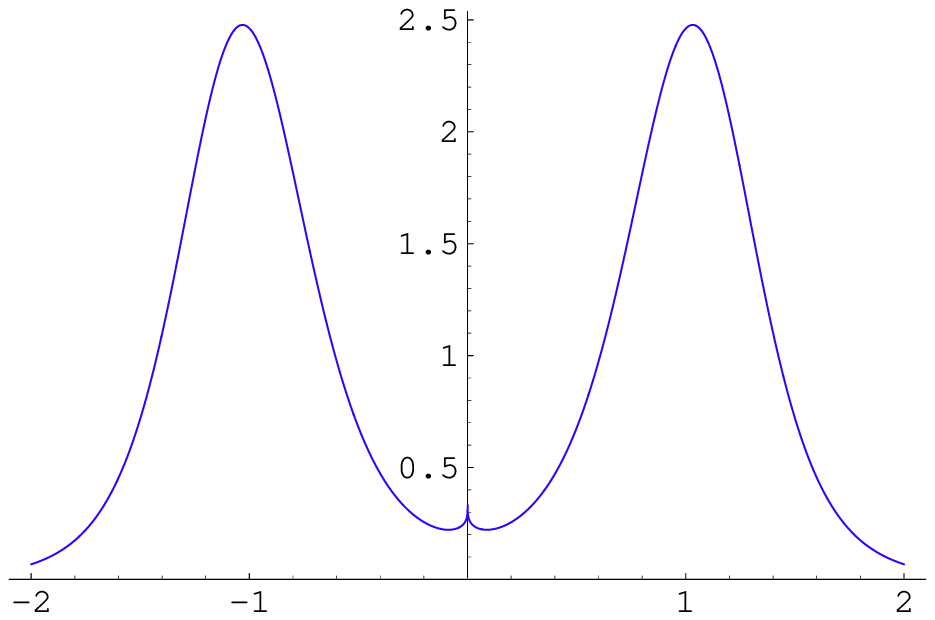} &
     \epsfxsize=1.5in
    \epsffile{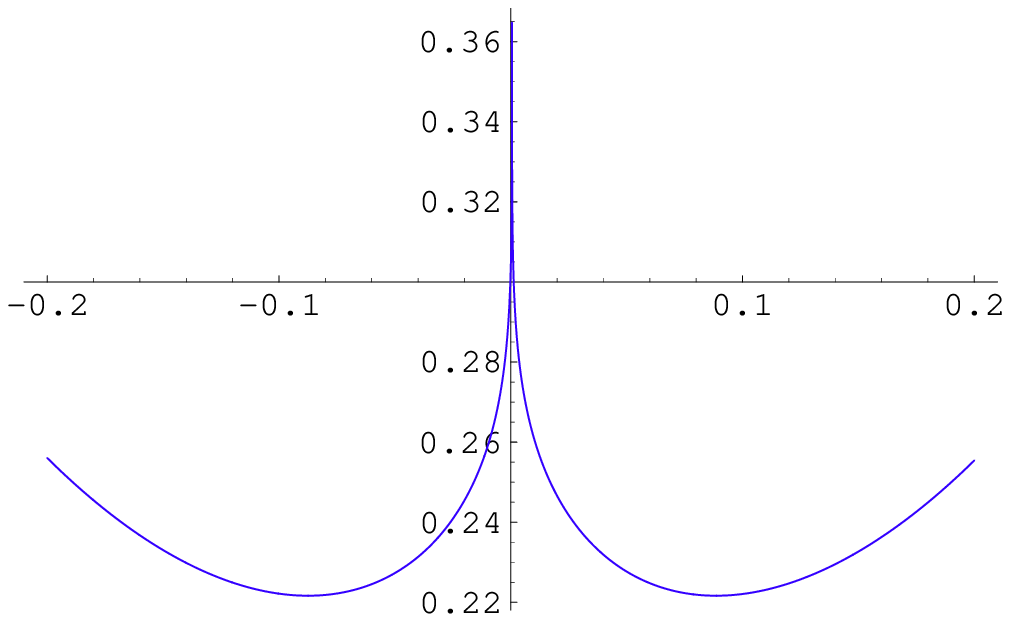}
\end{array}$
\end{center}
\caption{\footnotesize
Wall for the K\"{a}hler potential $h_2$
 and intermediate $l$ with respect to the previous figures.
 Left panel:
Field profiles for $\phi_s(z)$ and $\phi_\chi(z)$. Central panel: Energy density. Right panel: Zoom of the central region which
shows the presence of another energy peak at the center.}
\label{fig-surprise}
\end{figure}

A solution of the wall equation interpolating between the vacua at
$\phi_s=\pm \phi_{s0}$ can pass through the valley of vacua at
$\phi_s=0$; so it can be thought as composite of two walls, one
interpolating between $\phi_s=\phi_{s0}$ and $\phi_s=0$ and the
other interpolating
 $\phi_s=0$ and $\phi_s=-\phi_{s0}$. Hence, there is a hidden
modulus which describes the distance $l$ between the two component
walls. The parameter $l$
 can be defined for example as the distance in the $z$ axis between
the two peaks in the energy density. For large $l$ the wall has  a
three layers structure shown in Fig. \ref{fig2}: in the first layer
$\phi_s=\phi_{s0}$ jumps to zero while $\phi_\chi=\phi_{\chi 0}$ is
almost constant; in the second layer $\phi_s=0$ and $\phi_\chi$ goes
to the value $\phi_\chi=-\phi_{\chi 0}$; in the third layer $\phi_s$
jumps to the value of $\phi_s=-\phi_{s0}$. A more careful analysis
of the energy density profile reveals {\it always} the presence of a
third peak in the energy density
 just in between the other two symmetric walls
 in correspondence of the origin $\phi_s,\phi_\chi=0$
 (see  Fig. \ref{fig-surprise}).
The energy density evaluated at the origin diverges\footnote{
 The energy density of the BPS wall is given by the following expression:
 \beq \rho=2\left| \frac{d}{dz} Re(W)\right|=2 \left|\frac{d \phi_s}{dz}
 \frac{\partial W}{\partial \phi_s}\right|+
 2 \left|\frac{d \phi_\chi}{dz} \frac{\partial W}{\partial \phi_\chi}\right|. \eeq
We also have near the origin at $\phi_s,\phi_\chi=0$:
\beq  \frac{\partial W}{\partial \phi_\chi} \approx  \frac{\phi_s}
{\phi_\chi \log \phi_\chi^2}, \,\,\,\,
 \frac{\partial W}{\partial \phi_s} \approx \log \left(
  \frac{\phi_s}
{\phi_\chi \log \phi_\chi^2} \right). \eeq The only way one
can avoid a divergence  is by requiring that $ \frac{\phi_s} {\phi_\chi \log \phi_\chi^2}$ approaches a
non-zero constant. This does not happen here since $\phi_s \approx C
\phi_\chi$.}. Note that the superpotential is regular near the
origin.

 \begin{figure}[h]
\begin{center}
$\begin{array}{c@{\hspace{.2in}}c@{\hspace{.2in}}c} \epsfxsize=1.5in
\epsffile{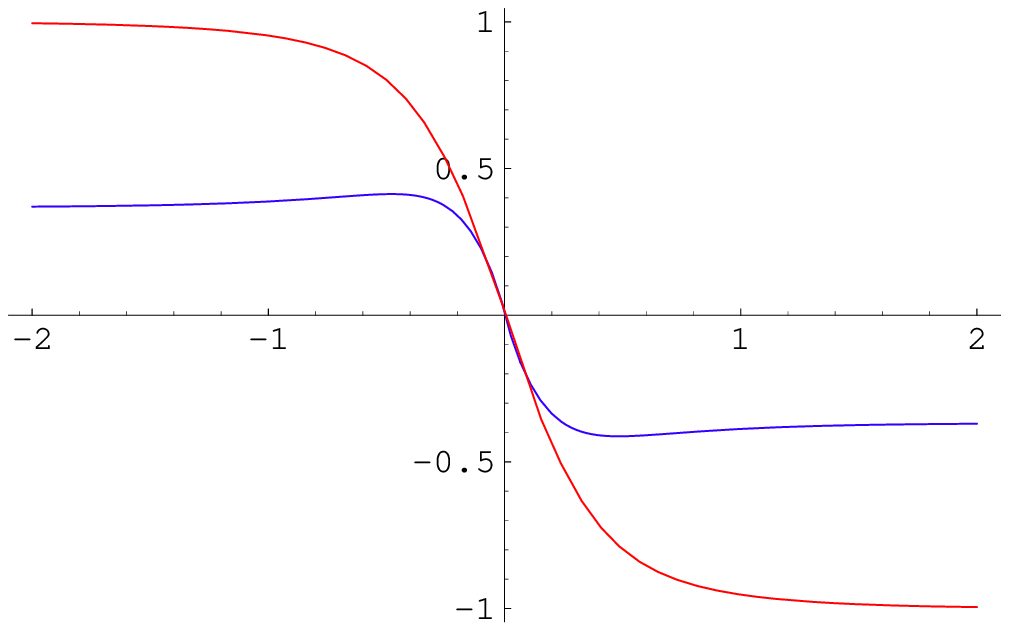} &
    \epsfxsize=1.5in
    \epsffile{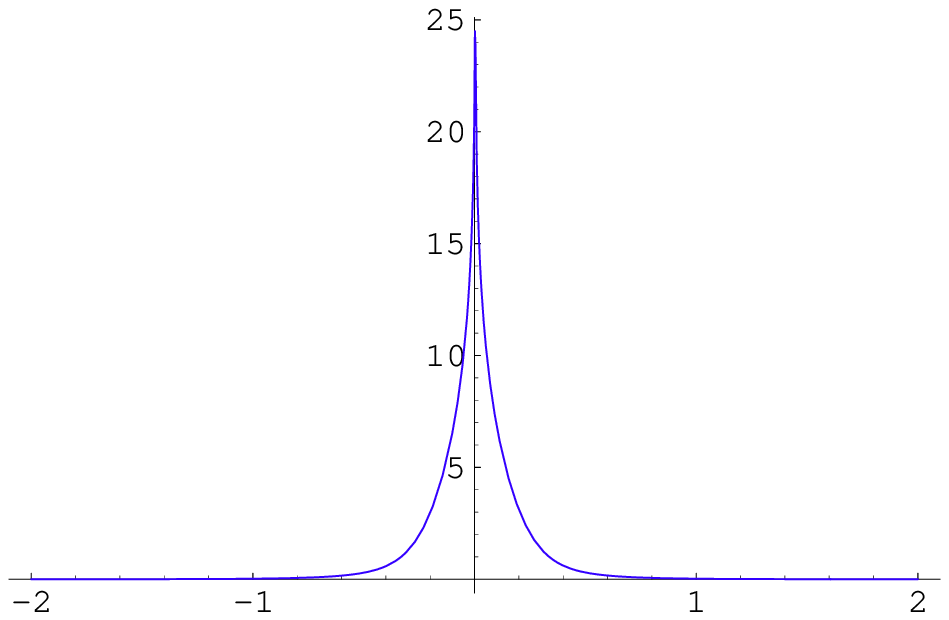} &
     \epsfxsize=1.5in
    \epsffile{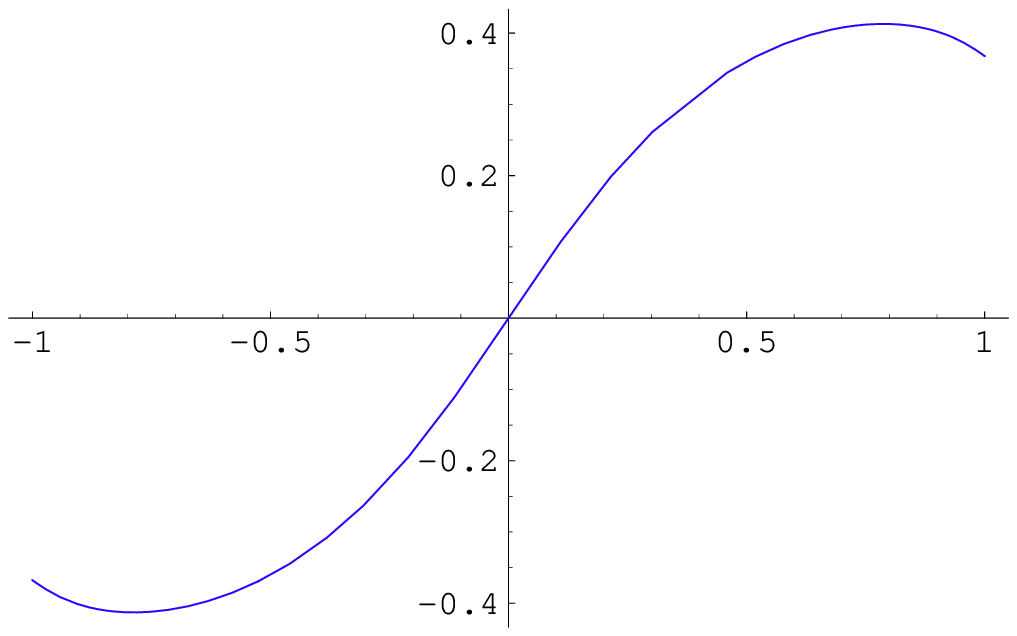}
\end{array}$
\end{center}
\caption{\footnotesize Wall for the K\"{a}hler potential $h_2$ and
with $l$ small enough that the domain walls fuse in a single one.
 Left:
Field profiles for $\phi_s(z)$ and $\phi_\chi(z)$.
Center: Energy density. Right: Field profiles in the plane $(\phi_s,\phi_\chi)$.}
\label{fig2bis}
\end{figure}

The distance $l$  is a good parameter for the moduli space only in
the limit in which the relative displacement of the two symmetric
peaks of the energy density with respect to the origin is large. As
can be seen in Fig \ref{fig2bis}, the two symmetric peaks in the
energy density merge with the one at the origin
 in the limit of a small relative distance.
A more precise way to parameterize the solutions is given by the
real and positive constant $C$ introduced in (\ref{mocambo}): \beq
C=\lim_{\phi_s,\phi_\chi \rightarrow 0}
 \frac{\phi_s}{\phi_\chi} \ .\eeq
 In the limit $C \rightarrow 0$ the distance $l$
  goes to infinity. For
  $C =\mathcal{O}(1)$ the two peaks in the energy density
  join with the third peak at the origin.

\begin{figure}[h]
\begin{center}
\epsfxsize=2.5in \epsffile{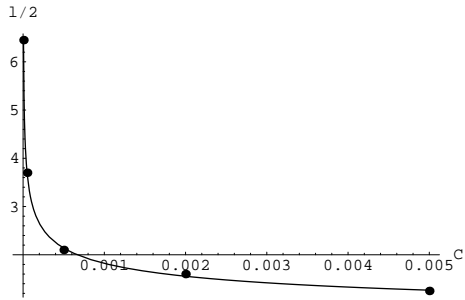}
\end{center}
\caption{\footnotesize Fit providing the behavior of $l/2$ as
function of $C$. The solid curve corresponds to the fitted curve
\ref{fitc} reported in the main text. The dots are some
representative points. } \label{figure-fit}
\end{figure}

In order to have a better understanding of our results and provide a
more precise relation between the hidden modulus $l$ and $C$ we have
performed a numerical fit (see Fig.~\ref{figure-fit}) which yields
\begin{eqnarray}
\frac{l\left(C\right)}{2}=(0.14 \pm 0.03) + \frac{0.298\pm
0.002}{C^{0.25}} \ ,
 \label{fitc}
\end{eqnarray}
where the errors are statistical in nature.

The situation looks similar to another model discussed in Ref.
\cite{muridegeneri,muridegeneri2}: this is a two superfield model with
superpotential: \beq W(\Phi_1,\Phi_2)=\frac{m^2}{\lambda} \Phi_1 -
\frac{1}{3} \lambda \Phi_1^3 - \alpha \Phi_1 \Phi_2^2\eeq and a
standard K\"{a}hler potential: \beq K=\Phi_1 \bar{\Phi}_1 + \Phi_2
\bar{\Phi}_2.\eeq There are four degenerate vacua in this model:
\beq \phi_1=\pm \frac{m}{\lambda}, \,\,\, \phi_2=0 \quad {\rm and}
\quad\phi_1=0, \,\,\, \phi_2=\pm \frac{m}{\sqrt{\lambda \alpha}}.
\eeq A family of degenerate BPS domain walls exists interpolating
the two vacua with $\phi_2=0$: this is due to the fact that this
wall can be thought as composite of two walls interpolating between
the $\phi_2=0$ vacua and the $\phi_1=0$ one. The relative distance
of the two elementary and symmetric walls with respect to the origin
is a modulus of this family of degenerate walls. There are similar
composite walls also in $\mathcal{N}=2$ $U(N)$  SQCD in the Higgs
phase  with a Fayet-Iliopolous term (see for example \cite{tong}).

The walls in the extended VY are somehow
different from the ones just mentioned  above. In this case
the solution passes through an entire valley of vacua at $\phi_s=0$
and not through just another vacuum state.

\section{Issue of complex walls}

Let us take $N=2$; if the field $S$ is integrated out, the following
result for $W_{eff}(\chi)$ is found: \beq W_{eff}(\chi)=\frac{4}{3}
e \chi \log \chi^2.\eeq It is easy to see that it is impossible to
avoid the log-cuts in the superpotential: there are cuts on the
imaginary positive and negative axis, and the only way to avoid them
is to pass through the origin. This problem is still present in the
complete superpotential: \beq W(S,\chi)=\frac{8}{3} S \left[
\log_{(0)} \left( \frac{S} { -e \frac{\chi}{2} \log_{(0)} \chi^2
}\right) -1  \right]. \eeq  Here too, the only way to avoid the cuts
on the positive and negative imaginary axis is to pass through
$\phi_\chi=0$;
 but if we do this keeping $\phi_s \neq 0$
 the superpotential diverges. This argument is not a rigorous
 non-existence proof, but shows that there is no obvious
 way to find such solutions.

In the framework of the Taylor, Veneziano and Yankielowicz
theory (\cite{TVY}), which is supposed to be the effective theory
for $\mathcal{N}=1$ $U(N)$ SQCD, this problem \cite{kss,smilga,dcm1} does not appear.
For example, in the two color and two flavor case (i.e. $N=2$, $N_f=1$) we have:
\beq W(S,M)=\frac{2}{3} S \log \frac{S M}{e \Lambda^5}+\frac{m}{2} M. \eeq
Here $M$ is the mesonic degree of freedom. Note also that here the normalization of the superpotential is slightly different than the one used
earlier.
With such a superpotential it is possible to find solutions
in which the phase of $S$ and the one of $M$ cancel each other
in such a way that the branch cuts of the logarithm are avoided.
This mechanism does not work in the extended VY theory, because of the presence of another
logarithm inside the argument of the first logarithm.

\section{Conclusions}
\label{conclu} In this work we have investigated the domain wall
problem related to the extended VY theory proposed in Ref.
\cite{EVY1}. No BPS domain wall solution has been found for the
general  case of complex walls. However real solutions exist
\cite{EVY3} for an even number of colors connecting two vacua
aligned with the origin. Surprisingly we have discovered the
presence of a new hidden noncompact modulus for the real walls.
Although the presence of an hidden modulus is interesting per se,
unfortunately it does not match the  expected spectrum of domain
walls for $\mathcal{N}=1$ super Yang-Mills. As discussed in Ref.
\cite{AV,RSV,ritz}, a discrete spectrum of wall with multiplicity
 \beq n_k=\frac{N !}{k ! (N-k)!} \label{molteplicita} \eeq
is expected. No theorem forbids a description of the SYM walls via
an effective theory of its composite states. An argument against
this possibility was suggested in Ref. \cite{witten} in the large
$N$ limit; however, as pointed out in Ref. \cite{dg}, this argument
can be avoided
 if the wall
thickness scales as $1/N$ at large $N$. Various models, more or less
linked to SYM were also considered in the past to mimic such a
behavior \cite{dg,dcm2}. A satisfactory effective description
 of these objects should explain the  discrete spectrum in Eq. (\ref{molteplicita}) while using
 fields directly linked to SYM \footnote{In Ref. \cite{ksy} domain walls were studied in the softly broken $\mathcal{N}=2$ theory. The breaking to $\mathcal{N}=1$ was achieved by adding a mass  term for
the multiplet in the adjoint representation. In this model the
picture proposed to understand the domain wall structure is indeed a
complex one; the domain wall connects a vacuum with a monopole
condensate to one with a dyonic condensate. Only one wall is found
and not two as expected; a conjecture in order to solve this problem
is presented in Ref. \cite{sy}. }. We have shown that although the
extended VY theory displays some domain wall solutions these are not
the ones expected on theoretical grounds.

At this point one may argue that the physics of the SYM walls should
be described not via confined objects such as gluino-balls or
glue-balls but via a more subtle substructure.

\section*{Acknowledgments}

We are grateful to P. Merlatti, M. Shifman, A. Vainshtein  and F. Vian for discussions and comments.

The work  of R.A.  is
supported in part by DOE grant DE-FG02-94ER408. The work of F.S. is supported by the Marie Curie Excellence Grant as team leader under contract MEXT-CT-2004-013510 and by the Danish Research Agency. F.S. thanks the William I. Fine Theoretical Physics Institute for the kind hospitality during the initial stages of this work.


\end{document}